\documentclass[runningheads]{llncs}
\usepackage[T1]{fontenc}
\usepackage{graphicx}
\usepackage{hyperref}
\usepackage{academicons}
\usepackage{amssymb}
\usepackage{xcolor}
\usepackage{float}
\usepackage{pgf-pie}
\usepackage{orcidlink}
\usepackage{amsmath}
\usepackage{booktabs}
\usepackage{colortbl}
\usepackage{tikz}
\usepackage{cite}
\usepackage{enumitem}
\usepackage{subcaption}
\usepackage[export]{adjustbox}
\usepackage[T1]{fontenc}
\usepackage{pgfplots}

\pgfplotsset{compat=1.17}
\usetikzlibrary{shapes, arrows}

\tikzstyle{block} = [rectangle, draw, fill=blue!20, 
    text centered, rounded corners, minimum height=2em, minimum width=4em]
\tikzstyle{arrow} = [thick,->,>=stealth]

\begin{document}

\title{Feature-level Site Leakage Reduction for Cross-Hospital Chest X-ray Transfer via Self-Supervised Learning\thanks{Accepted at The 7th International Conference on Computing Systems and Applications.}}

\titlerunning{ SSL Feature-level Site Leakage Reduction for Cross-Hospital Chest X-ray}

\author{Ayoub Bouaziz\inst{1,4}\orcidlink{0009-0008-9074-0374} \and
 Lokmane Chebouba\inst{1,2,3}\orcidlink{0000-0002-2131-6421}}

\authorrunning{Bouaziz and Chebouba}

\institute{University of Mentouri Brothers Constantine 1, Constantine, Algeria
\and LRIA Laboratory, University of Science and Technology Houari Boumediene (USTHB), Bab-Ezzouar, Algeria
\and LSIACIO Laboratory, University of Mentouri Brothers Constantine 1, Constantine, Algeria
\and University of Paris-VIII-Vincennes-Saint-Denis, France
\email{ayoublouaye.bouaziz@student.umc.edu.dz},\\
\email{lokmane.chebouba@umc.edu.dz},\\
\email{ayoub.bouaziz09@etud.univ-paris8.fr}}

\maketitle
\begin{abstract}
Cross-hospital failure in chest X-ray models is often attributed to domain shift, yet most work assumes invariance without measuring it. This paper studies how to measure site leakage directly and how that measurement changes conclusions about transfer methods.

We study multi-site self-supervised learning (SSL) and feature-level adversarial site confusion for cross-hospital transfer.
We pretrain a ResNet-18 on NIH and CheXpert without pathology labels.
We then freeze the encoder and train a linear pneumonia classifier on NIH only, evaluating transfer to RSNA.
We quantify site leakage using a post hoc linear probe that predicts acquisition site from frozen backbone features $f$ and projection features $z$.

Across 3 random seeds, multi-site SSL improves RSNA AUC from 0.6736 $\pm$ 0.0148 (ImageNet initialization) to 0.7804 $\pm$ 0.0197.
Adding adversarial site confusion on $f$ reduces measured leakage but does not reliably improve AUC and increases variance.
On $f$, site probe accuracy drops from 0.9890 $\pm$ 0.0021 (SSL-only) to 0.8504 $\pm$ 0.0051 (CanonicalF), where chance is 0.50.
On $z$, probe accuracy drops from 0.8912 $\pm$ 0.0092 to 0.7810 $\pm$ 0.0250.
These results show that measuring leakage changes how transfer methods should be interpreted: multi-site SSL drives transfer, while adversarial confusion exposes the limits of invariance assumptions.
\end{abstract}

\section{Introduction}
Cross-hospital generalization is a deployment risk in medical imaging.
Hospitals differ in scanners, protocols, compression, and preprocessing.
They also differ in cohort mix and labeling practices.
These differences create site-specific signals that can correlate with labels in the training data.
A model can learn shortcuts and then fail after deployment to a new site.

Two common response families exist.
Pixel-level harmonization tries to normalize appearance across sites.
Supervised domain adaptation tries to align domains using labeled target data.
Pixel-level methods can remove clinically relevant cues.
Supervised adaptation often needs labeled target data, which is unavailable in many deployments.

We study representation learning with unlabeled multi-site data.
We pretrain an encoder with contrastive self-supervised learning (SSL) on multiple sites.
We then freeze the encoder and train a linear pneumonia classifier on one labeled source site.
We evaluate on other sites that are unseen during supervised training.

We focus on measurement, not assumptions.
Many papers claim ``site invariance'' but do not quantify it.
We measure site leakage directly with post hoc probes that predict site from frozen embeddings.
We report probe accuracy on backbone features $f$ used by the downstream classifier, and on projection features $z$ used only for SSL.
Chance is 0.50 because each probe is binary in our leave-one-site-out settings.

We also avoid a one-split story.
We report transfer in both directions between NIH and CheXpert.
We also run leave-one-site-out across three sites (NIH, CheXpert, RSNA).
In this setting, we pretrain SSL on two sites and evaluate transfer to the held-out site.
We report mean and standard deviation over 3 seeds for both AUC and leakage probes.
In our leave-one-site-out runs, multi-site SSL improves transfer AUC by about 0.095.
Adversarial site confusion reduces measured leakage and can improve AUC under some shifts, but it does not guarantee gains.

\paragraph{Contributions.}
\begin{itemize}
\item A leakage measurement protocol that quantifies site information in representations using post hoc probes and multi-split evaluation. This protocol is the primary contribution of the paper.
\item A multi-site SSL baseline for chest X-rays that improves cross-site transfer when used as a frozen backbone, including leave-one-site-out results across three sites.
\item A feature-level adversarial variant (CanonicalF) that applies a site-adversarial head to backbone features $f$ that the downstream model uses.

\item A multi-split evaluation that includes two transfer directions (NIH$\rightarrow$others and CheXpert$\rightarrow$others) and leave-one-site-out over NIH, CheXpert, and RSNA, all reported with mean $\pm$ standard deviation over 3 seeds.
\item An empirical finding that multi-site SSL is the main driver of transfer in our setting, while adversarial confusion is a leakage reduction tool whose downstream benefit depends on the shift and can increase variance.
\end{itemize}

\section{Related Work}
\subsection{Self-supervised learning for medical imaging}
Contrastive SSL methods learn transferable features from unlabeled images via augmentation-based view agreement.
In medical imaging, SSL is used to reduce label dependence and improve transfer, especially in radiographs~\cite{lin2025self, ali2025multimodal, huang2025prototype, alsaafin2025marblix, abbas2025umamba, usharani2026advancements, wu2026mix, janarthanam2025self}.

\subsection{Domain adversarial learning}
Domain adversarial learning uses a domain classifier and gradient reversal to encourage domain confusion.
This is widely used in domain adaptation.
In medical imaging, it is often applied to scanners or hospitals, but claims of invariance are frequently not supported by explicit leakage measurements~\cite{zhang2025hyperspherical, li2025adaptive, zhang2026pathology, huang2025adaptive, wang2025blackbox, shin2025zaxis, takemoto2025vulnerability, nalajala2025brain}.

\subsection{Harmonization}
Harmonization methods try to reduce site effects at the pixel or feature level.
Pixel-level approaches risk changing clinical content.
Feature-level approaches require careful validation because removing site information can also remove label signal when label and site are correlated~\cite{tobari2025domain}.

\section{Method}

\subsection{Problem setting}
Each image $x$ has a pathology label $y$ for supervised training and a site label $s \in \{0,1\}$ that indicates acquisition source.
We aim to learn an encoder $f_\theta$ that supports cross-site transfer when frozen, and to quantify how much site information remains in its representations~\cite{zech2018variable, castro2020causes}.

We denote backbone features by
\[
h = f_\theta(x) \in \mathbb{R}^{512},
\]
and projection features by
\[
z = \mathrm{norm}(p_\psi(h)) \in \mathbb{R}^{d},
\]
where $p_\psi$ is a projection head and $\mathrm{norm}(\cdot)$ is $\ell_2$ normalization.

\subsection{Multi-site contrastive pretraining}
We pretrain $f_\theta$ and $p_\psi$ on unlabeled images from NIH (site 0) and CheXpert (site 1) using a SimCLR-style contrastive objective~\cite{chen2020simclr, azizi2021bigssl, chowdhury2021self}.
For each image, we sample two stochastic augmentations $(x_1,x_2)$ and compute:
\[
h_1 = f_\theta(x_1), \quad h_2 = f_\theta(x_2), \quad
z_1 = \mathrm{norm}(p_\psi(h_1)), \quad z_2 = \mathrm{norm}(p_\psi(h_2)).
\]
We minimize an InfoNCE loss $\mathcal{L}_{\text{inv}}(z_1,z_2)$ with temperature $\tau$.

\subsection{CanonicalF: adversarial site confusion on backbone features}
We use adversarial site confusion as a controlled stress test to examine whether reducing measurable site information improves transfer. Applying the adversary on backbone features lets us test this assumption directly on the features used by the downstream classifier.
We attach a site classifier $g_\phi$ to $h$ and train with gradient reversal~\cite{ganin2016dann, guan2021domain, zhang2026pathology}:
\[
\mathcal{L} = \mathcal{L}_{\text{inv}} + w_{\text{site}}\mathcal{L}_{\text{site}},
\]
where $\mathcal{L}_{\text{site}}$ is cross-entropy for predicting $s$ from a gradient-reversed copy of $h$.

We set $w_{\text{site}}$ using a small sweep on one seed with a proxy selection criterion that trades off target-side validation AUC and leakage on $h$, then fix $w_{\text{site}}$ for all reported seeds and splits.

\subsection{Downstream evaluation with frozen backbones}
We evaluate representation quality by freezing the backbone $f_\theta$ and training a linear pneumonia head on top of $h$~\cite{he2020moco, azizi2021bigssl}.
We report two transfer directions.

\textbf{NIH source.}
We train on NIH train labels, select by NIH validation AUC, and report AUC on NIH validation, RSNA, and CheXpert validation.

\textbf{CheXpert source.}
We train on CheXpert train labels, select by CheXpert validation AUC, and report AUC on CheXpert validation, RSNA, and NIH validation.

All results are reported as mean and standard deviation over three random seeds.

\subsection{Site leakage measurement}
We quantify site leakage using post hoc linear probes on frozen embeddings. Each probe is trained on a balanced sample of 7000 images per site  with fixed train test splits shared across methods.~\cite{raghu2021vision, snoek2022label}.
For each representation type ($h$ and $z$), we:
\begin{itemize}
\item sample a balanced set of images per site and per split
\item extract frozen embeddings
\item train multinomial logistic regression on a train split
\item report site classification accuracy on a held-out test split
\end{itemize}
Chance accuracy is $0.50$ for two-site probes.

We report leakage for ImageNet initialization before SSL, SSL-only after pretraining, CanonicalF after pretraining, and after supervised fine-tuning on the source labels to test whether leakage reduction persists under adaptation.
We also report a three-site probe (NIH, CheXpert, RSNA) where chance accuracy is $1/3$.
\begin{figure}[t]
\centering
\includegraphics[width=\textwidth]{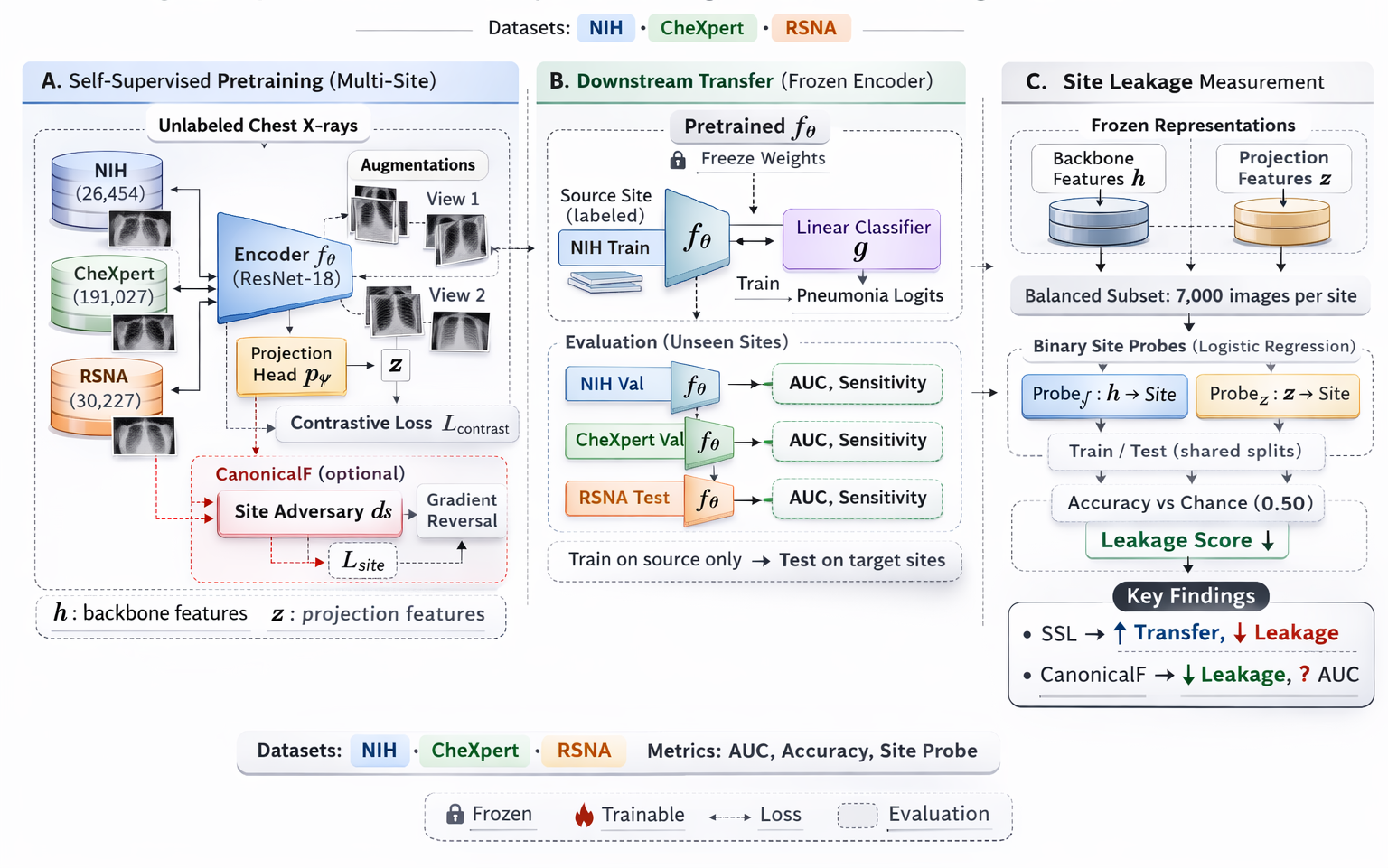}
\caption{Overview. (A) Multi-site SSL pretraining on NIH and CheXpert with balanced batches. (B) CanonicalF adds an optional site-adversarial head acting on backbone features $f$. (C) Downstream pneumonia detection trains a linear head on NIH with a frozen encoder. Site leakage is measured post hoc with logistic-regression probes on frozen $f$ and $z$ (chance = 0.50 for two sites).}
\label{fig:pipeline}
\end{figure}

\section{Experiments}
\subsection{Datasets}
We use three public chest X-ray datasets. 
NIH ChestX-ray14\footnote{\url{https://nihcc.app.box.com/v/ChestXray-NIHCC}}~\cite{wang2017chestxray14} provides labels for pneumonia and is used for supervised training and validation splits. 
CheXpert\footnote{\url{https://stanfordmlgroup.github.io/competitions/chexpert/}}~\cite{irvin2019chexpert} provides unlabeled images for SSL pretraining and a small labeled validation split used only for evaluation. 
RSNA Pneumonia Detection Challenge\footnote{\url{https://www.kaggle.com/c/rsna-pneumonia-detection-challenge}}~\cite{shih2019augmenting} is used only for evaluation as a held-out hospital.
\begin{table}[t]
\centering
\caption{Main distinct characteristics of the datasets used in this work.}
\label{tab:datasets_detailed}
\resizebox{\textwidth}{!}{%
\begin{tabular}{p{3.2cm} p{4.2cm} p{4.2cm} p{4.2cm}}
\toprule
\textbf{Characteristic} & \textbf{NIH ChestX-ray14} & \textbf{CheXpert} & \textbf{RSNA Pneumonia Challenge} \\
\midrule
Origin & NIH Clinical Center, USA & Stanford Hospital, USA & RSNA Pneumonia Detection Challenge (compiled from multiple institutions) \\
\addlinespace

Modality & Chest X-ray (2D) & Chest X-ray (2D) & Chest X-ray (2D) \\
\addlinespace

Labeling & Image-level labels (includes pneumonia) & Labels extracted from radiology reports (CheXpert labeler); uncertainty present & Pneumonia annotations for challenge (used here for evaluation) \\
\addlinespace

Views used in this work & Frontal only & Frontal only & Frontal only (as provided in split) \\
\addlinespace

Role in this paper & Supervised source site for linear head; also part of SSL pool & Unlabeled SSL pretraining site; extra target evaluation on CheXpert-valid & Held-out target evaluation for transfer (unseen during supervised training/selection) \\
\addlinespace

Size used in this work & NIH-train: 26,454 images; NIH-val: 2,940 images & Train: 191,027 images; Valid: 202 images (pneumonia task) & 30,227 images \\
\addlinespace

Class balance (pneumonia) & Train: 240 pos / 26,214 neg; Val: 27 pos / 2,913 neg & Train: 4,675 pos / 186,352 neg; Valid: 8 pos / 194 neg & 9,555 pos / 20,672 neg \\
\addlinespace

Availability & Public & Public (dataset license terms apply) & Public (competition dataset terms apply) \\
\bottomrule
\end{tabular}}
\end{table}
\subsection{Implementation details}
Backbone: ResNet-18 initialized from ImageNet, converted to single-channel input by averaging RGB weights.
SSL: InfoNCE with temperature $\tau = 0.2$, AdamW with learning rate $3 \times 10^{-4}$ and weight decay $10^{-4}$, 20 epochs, batch size 128.
We use balanced batches with half NIH and half CheXpert per batch.
Augmentations: random resized crop and small rotation, no horizontal flip.

CanonicalF: same SSL loss plus site loss on backbone features $f$, with gradient reversal.
We select $w_{\text{site}} = 0.2$ from a fast sweep on seed 0.

Downstream: frozen backbone and a linear head trained on NIH train for 20 epochs with AdamW learning rate $10^{-4}$ and weight decay $10^{-4}$.
Because NIH pneumonia is highly imbalanced, we use BCEWithLogitsLoss with \texttt{pos\_weight} computed from NIH train (109.225).
\subsection{Evaluation protocols}
We report two downstream transfer directions and two leakage settings.

\textbf{Frozen-backbone transfer AUC.}
We freeze the encoder and train a linear pneumonia head on top of $h$.
We use a labeled source split for training and a source validation split for model selection.
We then evaluate AUC on unseen sites.

\begin{itemize}
\item \textbf{NIH source:} train on NIH train, select by NIH validation, evaluate on RSNA and CheXpert validation.
\item \textbf{CheXpert source:} train on CheXpert train, select by CheXpert validation, evaluate on RSNA and NIH validation.
\end{itemize}

\textbf{Leave-one-site-out transfer.}
For three-site leave-one-site-out, we hold out one site as the target.
We pretrain SSL on the remaining two sites (unlabeled), freeze the encoder, train the linear head on the labeled source site among the remaining sites, select on that source validation split, and evaluate on the held-out target site.

\textbf{Leakage probes.}
We report leakage with logistic-regression probes on frozen embeddings.
We report:
\begin{itemize}
\item \textbf{Binary probe (NIH vs CheXpert):} $K=2$, chance $=0.50$.
\item \textbf{Three-site probe (NIH vs CheXpert vs RSNA):} $K=3$, chance $=1/3$.
\end{itemize}

All reported AUC and probe accuracies use mean $\pm$ standard deviation over three random seeds.
\section{Results}
\subsection{Transfer performance with statistics}
Table \ref{tab:auc} reports mean and standard deviation over 3 seeds.

\begin{table}[t]
\centering
\caption{AUC (mean $\pm$ std) over 3 seeds. Training labels are NIH only. Backbones are frozen.}
\label{tab:auc}
\begin{tabular}{lccc}
\toprule
Method & NIH val AUC & RSNA AUC & CheXpert valid AUC \\
\midrule
ImageNet Frozen  & 0.5787 $\pm$ 0.0789 & 0.6736 $\pm$ 0.0148 & 0.5939 $\pm$ 0.0352 \\
SSL-only Frozen  & 0.6220 $\pm$ 0.0408 & 0.7804 $\pm$ 0.0197 & 0.7309 $\pm$ 0.0554 \\
CanonicalF Frozen & 0.5923 $\pm$ 0.0152 & 0.7241 $\pm$ 0.0956 & 0.7998 $\pm$ 0.0399 \\
\bottomrule
\end{tabular}
\end{table}

Key observations:
\begin{itemize}
\item Multi-site SSL improves RSNA transfer relative to ImageNet initialization.
\item CanonicalF does not consistently beat SSL-only on RSNA and shows high variance across seeds.
\item CanonicalF improves CheXpert validation AUC in these runs, but that split is small and should be treated as a weak signal.
\end{itemize}
\subsubsection{Stability of adversarial training:}

CanonicalF increases RSNA AUC variance relative to SSL-only. The coefficient of variation rises from roughly 0.025 to about 0.13, indicating that adversarial training introduces sensitivity to initialization and weight selection. This instability suggests that reducing site signal can conflict with preserving predictive signal, which limits reliability in deployment settings.
\subsection{Measured site leakage}
Table \ref{tab:leak} reports site probe accuracy (mean $\pm$ std) on frozen representations.
Chance is 0.50.
\begin{figure}[t]
\centering
\includegraphics[width=\textwidth]{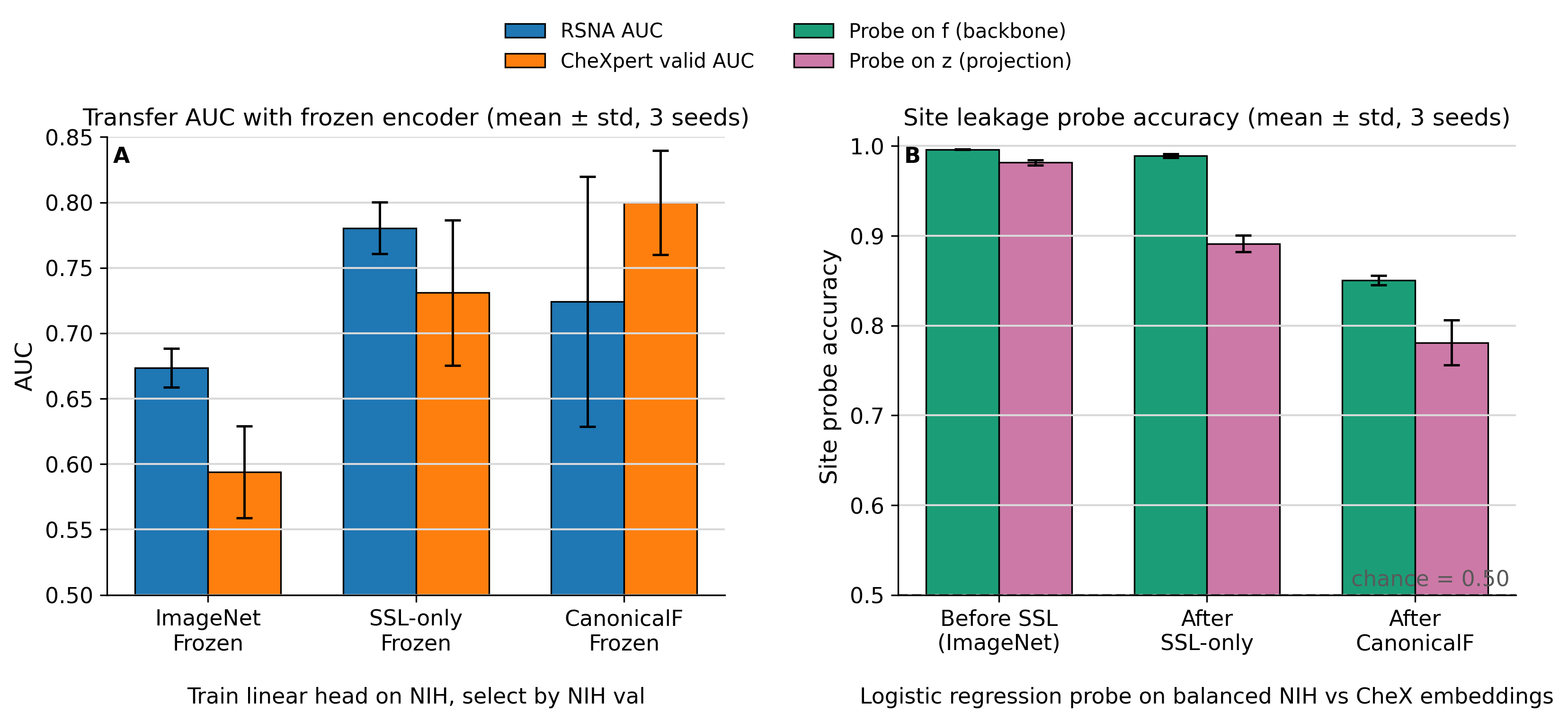}
\caption{Transfer performance and site leakage (mean $\pm$ std over 3 seeds). Left: AUC with a frozen encoder and a linear pneumonia head trained on NIH and selected by NIH validation. Right: site leakage probe accuracy (logistic regression on balanced NIH vs CheX embeddings) for backbone features $f$ and projection features $z$; chance = 0.50. Multi-site SSL improves transfer AUC, while CanonicalF reduces measured leakage but does not guarantee higher transfer AUC.}
\label{fig:results}
\end{figure}
\begin{table}[t]
\centering
\caption{Site leakage via linear probe accuracy (mean $\pm$ std) over 3 seeds. Chance is 0.50.}
\label{tab:leak}
\begin{tabular}{lcc}
\toprule
Stage & Probe on $f$ & Probe on $z$ \\
\midrule
Before SSL (ImageNet) & 0.9962 $\pm$ 0.0004 & 0.9815 $\pm$ 0.0030 \\
After SSL-only        & 0.9890 $\pm$ 0.0021 & 0.8912 $\pm$ 0.0092 \\
After CanonicalF      & 0.8504 $\pm$ 0.0051 & 0.7810 $\pm$ 0.0250 \\
\bottomrule
\end{tabular}
\end{table}

CanonicalF reduces measured leakage on both $f$ and $z$ relative to SSL-only.
However, leakage remains far above chance.
This does not support a claim of site-invariant representations.
It supports a claim of measurable leakage reduction.

\subsection{What the results imply for the main claim}
In this setting, the main driver of cross-hospital transfer is multi-site SSL.
Adversarial confusion reduces measured site leakage but does not guarantee better transfer AUC.
The correct claim is:
\begin{quote}
Multi-site SSL improves transfer, and feature-level adversarial confusion is an optional tool that can reduce site leakage and may help under some shifts.
\end{quote}

\section{Discussion}
\subsection{Why invariance is hard here}
The site probe on ImageNet features is near perfect.
This suggests strong, easily separable site cues in the raw data distribution.
A small adversary and short SSL schedule are unlikely to remove all such cues without also damaging predictive content.
Also, label and site can be correlated.
Pushing too hard toward confusion can remove signal useful for the downstream label.

\subsection{Limitations}
We use only two sites for SSL pretraining.
We evaluate on a limited set of transfer directions.
We use linear probing as the leakage metric, which is informative but not a full characterization of mutual information.
CheXpert validation is small in our split and should not be over-interpreted.
We do not claim invariance, only leakage reduction as measured by the probe.

\subsection{Next steps}
Future work can expand to stronger backbones and non-linear probes, but our current study focuses on explicit leakage measurement under frozen-encoder transfer. We plan to add several Databases for SSL pretraining To achieve invariance.

\section{Conclusion}
We evaluated multi-site SSL and feature-level adversarial site confusion for chest X-ray transfer.
Multi-site SSL improves RSNA transfer when the backbone is frozen.
Feature-level adversarial training reduces measured site leakage on both backbone and projection features, but leakage remains far above chance and AUC gains are not consistent.
These results motivate reporting leakage metrics alongside transfer performance, and they support leakage reduction claims rather than invariance claims.
\section*{Acknowledgments}
The authors acknowledge the use of AI-assisted tools exclusively for improving the clarity and grammar of the manuscript. 
No part of the scientific content, experimental design, data analysis, or code was generated by artificial intelligence systems. 
All intellectual contributions are the authors' original work.

\end{document}